\documentclass[aps,prl,preprint,amsmath,amssymb]{revtex4-1}
\usepackage{graphicx}
\usepackage{epstopdf}
\usepackage{longtable}
\usepackage{hyperref}
\usepackage{amssymb}
\usepackage{dcolumn}
\usepackage{bm}% bold math
\usepackage{color}

\begin{document}

\title{Quantum Criticality in a Layered Iridate}

\author{Kousik Samanta}
\affiliation{``Gleb Wataghin'' Institute of Physics - IFGW, University of Campinas - UNICAMP, Campinas, Brazil}

\author{Jean C. Souza}
\affiliation{``Gleb Wataghin'' Institute of Physics - IFGW, University of Campinas - UNICAMP, Campinas, Brazil}

\author{Danilo Rigitano}
\affiliation{``Gleb Wataghin'' Institute of Physics - IFGW, University of Campinas - UNICAMP, Campinas, Brazil}

\author{Adimir I. Morales}
\affiliation{``Gleb Wataghin'' Institute of Physics - IFGW, University of Campinas - UNICAMP, Campinas, Brazil}

\author{Pascoal G. Pagliuso}
\affiliation{``Gleb Wataghin'' Institute of Physics - IFGW, University of Campinas - UNICAMP, Campinas, Brazil}

\author{Eduardo Granado}
\affiliation{``Gleb Wataghin'' Institute of Physics - IFGW, University of Campinas - UNICAMP, Campinas, Brazil}
\email{egranado@ifi.unicamp.br}

\begin{abstract}

Iridates provide a fertile ground to investigate correlated electrons in the presence of strong spin-orbit coupling. Bringing these systems to the proximity of a metal-insulator quantum phase transition is a challenge that must be met to access quantum critical fluctuations with charge and spin-orbital degrees of freedom. Here, electrical transport and Raman scattering measurements provide evidence that a metal-insulator quantum critical point is effectively reached in 5 \%\ Co-doped Sr$_2$IrO$_4$ with high structural quality. The {\it dc}-electrical conductivity shows a linear temperature dependence that is successfully captured by a model involving a Co acceptor level at the Fermi energy that becomes gradually populated at finite temperatures, creating thermally-activated holes in the $J_{\text {eff}}=1/2$ lower Hubbard band. The so-formed quantum critical fluctuations are exceptionally heavy and the resulting electronic continuum couples with an optical phonon at all temperatures. The magnetic order and pseudospin-phonon coupling are preserved under the Co doping. This work brings quantum phase transitions, iridates and heavy-fermion physics to the same arena.
\end{abstract}

\maketitle

\section{Introduction}

\begin{figure}[t]
\includegraphics[width=0.75\textwidth]{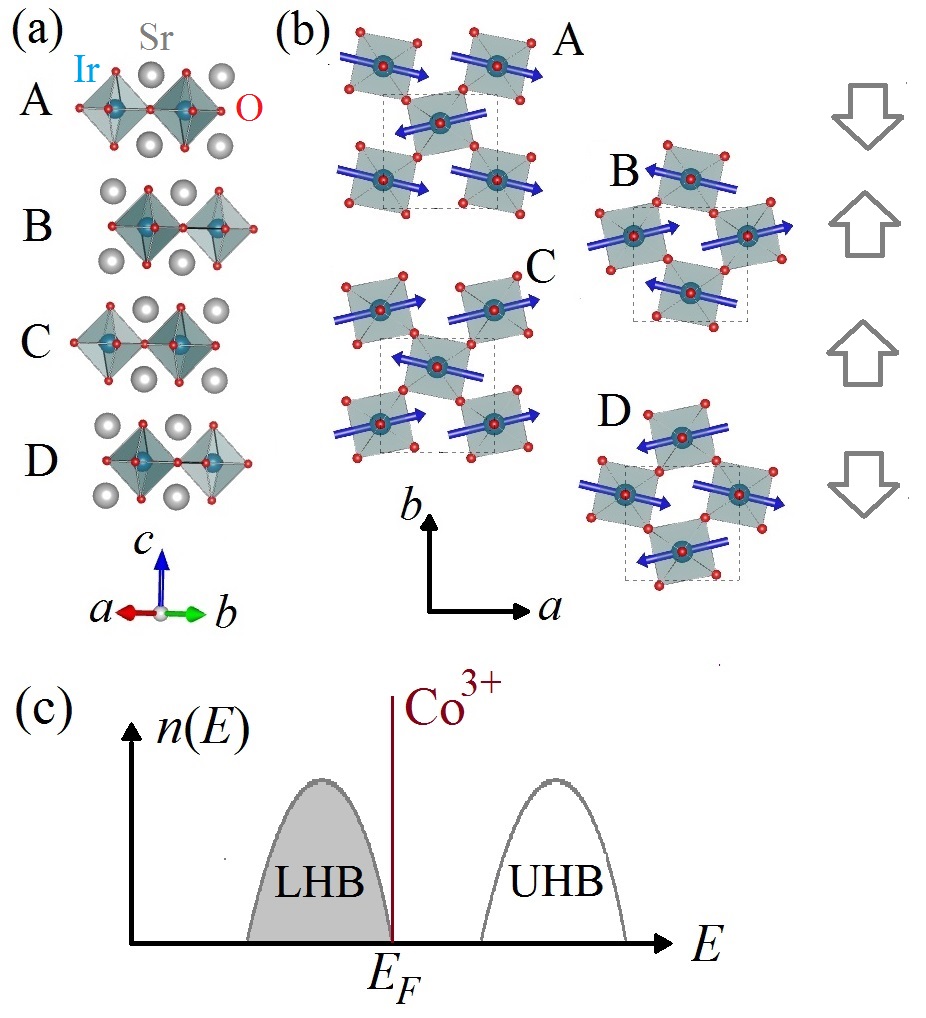}
\caption{\label{struc}  {\bf Crystal, magnetic and electronic structures.} (a) Crystal structure of Sr$_2$IrO$_4$ (SIO) \cite{Crawford1994,Huang1994} (A-D layers are identified). (b) Detailed view of each layer projected into the {\it ab} plane, showing the rotation pattern of the IrO$_6$ octahedra and the orientation of the Ir magnetic moments \cite{Kim2009,Ye2013,Boseggia2013}. The net moment of each layer is indicated in the right, leading to the $\downarrow \uparrow \uparrow \downarrow$ stacking pattern at zero magnetic field. (c) Schematic density of states $n(E)$ of Sr$_2$Ir$_{0.95}$Co$_{0.05}$O$_4$ (SICO) near the Fermi level $E_{\text F}$, showing the $J_{\text {eff}}=1/2$ lower an upper Hubbard bands (LHB and UHB) and a localized Co acceptor level. A metallic or Anderson insulating phase would arise if the energy $E_{\text Co^{3+}}$ of the acceptor level was located below $E_{\text F}$ (corresponding to a Co$^{3+}$ doping), whereas a gapped insulator would occur if instead this level was above $E_{\text F}$ (corresponding to a Co$^{4+}$ isoelectronic substitution). The quantum phase transition occurs where $E_{\text {Co}^{3+}} \sim E_{\text F}$, and the corresponding quantum critical fluctuations are electronic excitations from the Lower Hubbard band to the acceptor level.}
\end{figure}

A rare combination of strong spin-orbit coupling and substantial electron correlations in $5d^5$ iridates leads to exotic quantum states and remarkable physical phenomena such as spin liquid phases \cite{Okamoto2007}, Kitaev physics \cite{Jackeli2009,Price2012,Singh2012}, Fermi arcs in the electronic structure with a possible connection to high-$T_{\text c}$ superconductivity \cite{Wang2011,Watanabe2013,Meng2014,Kim2014,Yan2015,Kim2016}, and control of the crystal structure and magnetic properties by an electric current \cite{Cao2018electrical,Ye2020}. An intensively investigated iridate is Sr$_2$IrO$_4$ (SIO), crystallizing in a tetragonal structure with an $ABCD$ stacking of layers with tilted IrO$_6$ octahedra, as shown in Figs. \ref{struc}(a) and \ref{struc}(b) \cite{Crawford1994,Huang1994}. As for the electronic structure, the octahedral crystal field splits the Ir $5d$ level into $t_{\text {2g}}$ and $e_{\text g}$ levels. The strong spin-orbit coupling breaks the 6-fold degeneracy of the $t_{\text {2g}}$ levels, giving rise to lower $J_{\text {eff}} = 3/2$ and upper $J_{\text {eff}} = 1/2$ sublevels. Finally, the $J_{\text {eff}} = 1/2$-derived band is broken into lower and upper Hubbard bands, rendering SIO insulating \cite{Moon2008,Kim2009,Arita2012}. The Ir pseudospins order in a non-collinear magnetic structure below $T_{\text N} \sim 240$ K [see Fig. \ref{struc}(b)] \cite{Kim2009,Ye2013,Boseggia2013}. Injection of charge carriers through electron or hole doping tends to reduce $T_{\text N}$, although it is seems possible to destroy the insulating state without necessarily destroying the magnetic order \cite{Ge2011} and vice-versa \cite{Wang2015}. The most obvious approach to access the possible quantum critical fluctuations (QCF) \cite{Dobrosavljevic2012,Sachdev2008} associated with a metal-insulator transition would be to force the closure of the bandgap by application of external pressures. Nonetheless, the non-metallic state persists up to at least 185 GPa despite the identification of exotic states at intermediate pressures \cite{Haskel2012,Zocco2014,Samanta2018,Samanta2020,Haskel2020,Chen2020}. The alternative approach employed here involves the quest for a specific dilute cationic substitution that induces acceptor or donor impurity levels within the bandgap without actually charge doping the Ir-derived bands at $T=0$ K, as shown in Fig. \ref{struc}(c). A possible contender might be the Sr$_2$Ir$_{1-x}$Rh$_x$O$_4$ system, as Rh and Ir are in the same group of the periodic table. However, it has been demonstrated that dilute Rh-substitution induces Rh$^{3+}$/Ir$^{5+}$ charge partitioning, in practice doping the system with holes and rapidly suppressing the magnetic order \cite{Klein2008,Qi2012,Clancy2014,Ye2015,Brouet2015,Chikara2015,Cao2016,Sohn2016,Chikara2017,Louat2018,Xu2020,Zwartsenberg2020}. Thus, the energy of the Rh$^{3+}$ acceptor level is inferred to lay below the Fermi energy, $E_{\text {Rh}^{3+}} < E_{\text F}$,  at least for $x<0.24$. Another candidate is the Sr$_2$Ir$_{1-x}$Co$_x$O$_4$ system.  Based on first principle calculations for Sr$_2$Ir$_{0.5}$Co$_{0.5}$O$_4$, it was inferred that Co$^{3+}/$Ir$^{5+}$ charge partitioning is energetically stable \cite{Ou2014}, which was confirmed by subsequent experiments on this material \cite{Mikhailova2017,Chin2017,Agrestini2018}. On the other hand, more dilute Co substitutions have not been investigated in much detail yet, although it is already known that even moderate Co substitution levels ($x \leq 0.1$) are not sufficient to reduce $T_{\text N}$ appreciably \cite{Gatimo2012}. In this work, we investigate the charge transport, magnetic, structural and vibrational properties of Sr$_2$Ir$_{0.95}$Co$_{0.05}$O$_4$ (SICO). We demonstrate that this material shows charge transport quantum critical behavior that is consistent with a simple picture where the Co$^{3+}$ acceptor level coincides with the top of the lower Hubbard band at $E_{\text F}$ [see Fig. \ref{struc}(c)], triggering QCF with remarkably large renormalized masses.

\section{Results}

\subsection{Structural characterization}

X-ray powder diffraction profiles of SICO and SIO are shown for a selected angular interval in Figs. \ref{Raman1}(a) and \ref{Raman1}(b), respectively. The whole profiles are shown in the Supplementary Figures S1 and S2. Our data show asymmetric Bragg peak lineshapes in SICO [see Fig. \ref{Raman1}(b)], revealing stacking faults \cite{Rams2003}. This is likely caused by the random contribution to the interatomic elastic potencial caused by the Co substitution in SICO, disturbing the long-range ABCD stacking pattern of SIO [see Fig. \ref{struc}(a)]. The internal structural integrity of the layers is investigated by vibrational spectroscopy. Raman spectra of both samples are shown in Fig. \ref{Raman1}(c) at $T=20$ K. The main observed phonon modes, labelled as $M_1-M_4$ \cite{Samanta2018}, show similar positions and lineshapes for both compounds, being only slightly broader for SICO. An additional feature is observed at $\sim 340$ cm$^{-1}$ in SIO, indicative of a stoichiometric sample \cite{Sung2016,Glamazda2014}, whereas in SICO this feature is washed out and broad additional signals are observed at $\sim 260$ and $\sim 420$ cm$^{-1}$. The 260 cm$^{-1}$ peak was previously observed and becomes stronger in more disordered samples \cite{Sung2016,Glamazda2014}. The weak intensity of the additional Raman features associated with disorder and the relatively sharp $M_1-M_4$ modes in SICO indicate that the individual layer structures are preserved to a large extend through the 5 \%\ Co substitution, in contrast to the Sr$_2$Ir$_{1-x}$Ru$_x$O$_4$ series that shows much larger modifications in the phonon Raman spectra even at low doping levels \cite{Glamazda2014}.

\subsection{Charge transport at zero magnetic field}

The electrical resistivity curves $\rho(T)$ of SIO and SICO are shown in Fig. \ref{rho}(a). The values of $\rho(T)$ are significantly reduced for SICO with respect to the parent compound. In addition, the temperature dependence of the former is remarkably well captured by a simple power-law behavior, $\rho(T) = A^{-1} \cdot T^{-x}$ with $x=1.075(1)$, over the entire investigated temperature interval $3 < T < 300$ K [dashed line in Fig. \ref{rho}(a)].  Such power-law behavior is not found in SIO (see Supplementary Figure S3). Figure \ref{rho}(b) displays the same data in terms of electrical conductivity $\sigma(T)$ in a double-log scale for SICO only. The fit to the low-temperature data ($T<6$ K) is optimized by a slightly different exponent $x=1.003(6)$ that is even closer to unity. Note that the positive exponent implies that $\sigma \rightarrow 0$ as $T \rightarrow 0$, i.e., the material is non-metallic. On the other hand, an exponential behavior $\sigma(T) \propto \mathrm{exp}[-(T_0/T)^{\alpha}]$ is expected for insulating materials, with $1/4 \leq \alpha \leq 1$ for either a bandgap insulator ($\alpha = 1$) or for a variable range hopping mechanism where the specific $\alpha$ is defined by the system dimensionality and the specific energy-dependence of the density of states $n(E)$ near $E_{\text F}$ \cite{Mott1993}. Neither of such exponential behaviors for $\sigma(T)$ are observed for SICO (see Supplementary Figure S4), so this material is not classified either as a Mott, Slater or an Anderson insulator. Actually, according to very general scaling considerations, a power-law behavior $\sigma(T) \propto T^x$ signals a metal-insulator quantum phase transition \cite{Dobrosavljevic2012}. The positive exponent contrasts with the negative values found in metallic materials where the quantum criticality is not due to the proximity of a metal-insulator transition but interfaces two different conducting states, such as in a magnetic/heavy fermion quantum phase transition \cite{Schlottman2015,Paschen2020}. The specific value $x \sim +1$ for SICO must be captured by an appropriate microscopic model (see below).

\subsection{Magnetic properties and magnetoresistance}

\begin{figure}[t]
	\includegraphics[width=0.8\textwidth]{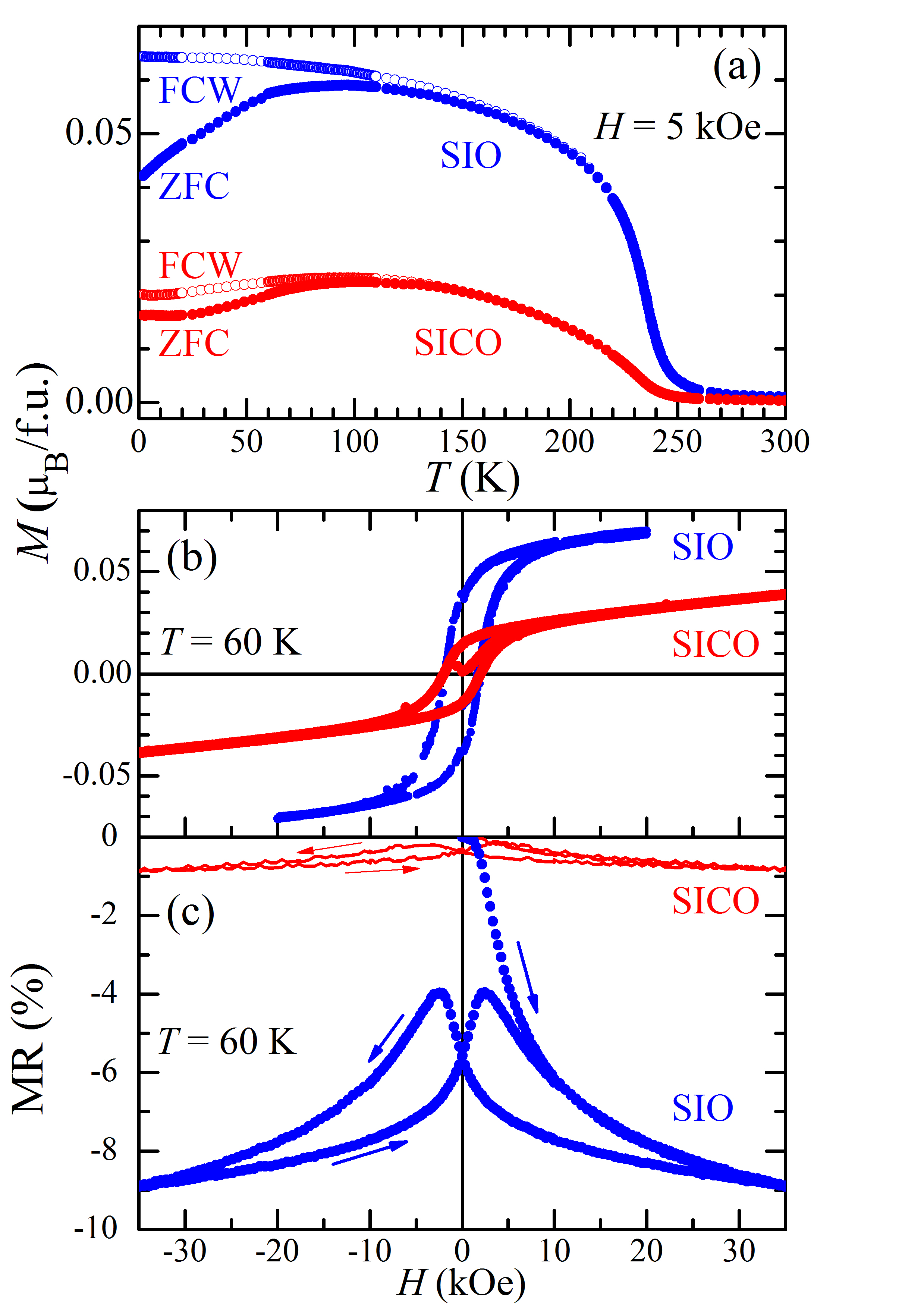}
\caption{\label{Mag} {\bf Magnetization and magnetoresistance.} (a) Temperature dependence of the magnetization $M(T)$ of  Sr$_2$IrO$_4$ (SIO) and Sr$_2$Ir$_{0.95}$Co$_{0.05}$O$_4$ (SICO) taken on warming with a magnetic field $H=5$ kOe that is applied after zero-field cooling (ZFC) or field cooling (FCW). $H$-dependence of magnetization (b) and magnetoresistance MR$\equiv [\rho(H)-\rho(H=0))/\rho(H=0)$] (c) for both materials at $T=60$ K, where $\rho(H)$ is the magnetic-field dependent electrical resistivity.} 
\end{figure}

Magnetization $M(T)$ curves of SICO and SIO are shown in Fig. \ref{Mag}(a), taken on warming after zero-field cooling and warming after field cooling, with $H=5$ kOe. Both compounds order at $T_N \sim 240$ K, in line with Ref. \cite{Gatimo2012}. The shapes of the $M(T)$ curves of SIO and SICO are distinct and the magnetization values are substantially reduced by the Co substitution. The latter result either indicates that the Co moments partially compensate the Ir net moments or the magnetic canting angle of the Ir moments [see Fig. \ref{struc}(b)] is reduced for SICO. Also, both materials feature a separation of the zero-field cooling and warming after zero-field cooling curves that becomes more prominent below $T \sim 100$ K \cite{Bhatti_2014}. This is understood in terms of a field-induced transition of the $\downarrow \uparrow \uparrow \downarrow$ magnetic structure and symmetry-related domains at zero-field to $\uparrow \uparrow \uparrow \uparrow$ under an applied field along the {\it ab}-plane $H_{\text {ab}} \sim 2$ kOe \cite{Kim2009,Sung2016,Porras2019}. In a polycrystal, the $H_{\text a}$ projection is different for each crystallite, and with an external field $H=5$ kOe a substantial fraction of the crystallites will still have $H_{\text a} < 2$ kOe, leading to a coexistence of different phases at low temperatures in proportions that are arguably sensitive to the thermomagnetic sample history. This is manifested not only in the different zero-field cooling and warming after zero-field cooling $M(T)$ curves, but also in the hysteresis of the isothermal $M(H)$ curves [see Fig. \ref{Mag}(b) and Ref. \onlinecite{Crawford1994}].

Figure \ref{Mag}(c) shows the magnetoresistance of SIO and SICO at $T=60$ K taken after zero-field cooling (see Supplementary Figure S5 for magnetoresistance at other temperatures). SIO shows a substantial negative magnetoresistance of $\sim 9$ \% for $H=35$ kOe. A field-hysteretic behavior is also observed. This phenomenon is again originated in the field-induced transition between the $\downarrow \uparrow \uparrow \downarrow$ and $\uparrow \uparrow \uparrow \uparrow$ structures, where the former is substantially more resistive \cite{Ge2011}. Remarkably, the initial high-resistance state is not recovered by cycling the field. This is likely due to a powder distribution of internal fields that are created after the $\uparrow \uparrow \uparrow \uparrow$ state is first activated, which may prevent the homogeneous high-resistance $\downarrow \uparrow \uparrow \downarrow$ state to be simultaneously recovered over the entire sample volume. The magnetoresistance of SICO is largely suppressed with respect to SIO [see Fig. \ref{Mag}(c)], remaining below 1 \% up to $H=35$ kOe. This indicates that, for SICO, the additional electronic scattering channel introduced by the Co impurities reduces the mean-free path of the thermally-activated charge carriers, overwhelming the magnetic scattering channel that is responsible for the large magnetoresistance of SIO. We therefore estimate that the electronic mean free path of SICO is $D \sim 8$ \AA, which is the average separation between the Co scattering centers for 5 \%\ substitution. We should note that in SIO the conductivity in the {\it ab} plane is higher than along {\it c} \cite{Cao1998,Ge2011}. Thus, in polycrystals the resistivity and magnetoresistance curves are presumably dominated by the charge transport in the {\it ab} plane. This is also likely valid for SICO, although this assumption is not essential for the conclusions of this work.

\subsection{Electron-phonon and pseudospin-phonon couplings}

We now return to a more detailed investigation of the Raman spectra [Fig. \ref{Raman1}(c) and Supplementary Figure S6]. The $M_1$ lineshape of SICO is fitted to the Fano expression $I(\omega)=I_0(q+\epsilon)^2/(1+\epsilon^2)$, where $I_0$ is the intensity, $\epsilon=(\omega-\omega_0)/\Gamma$, $\omega_0$ and $\Gamma$ are the phonon frequency and linewidth, respectively, and $1/q$ is the asymmetry parameter that measures the coupling of the phonons with the electronic continuum \cite{Fano1961} [see Fig. \ref{Raman2}(a)]. In opposition to SICO, the $M_1$ lineshape for SIO is symmetric at low temperatures, in agreement with previous studies \cite{Gretarsson2016,Cetin2012}.  Figure \ref{Raman2}(b) shows the temperature dependence of $|1/q|$ for mode $M_1$ of SICO. A substantial asymmetry parameter $|1/q|=0.15$ is observed at the base temperature. This reveals a continuum of low-energy excitations at $T \rightarrow 0$ that is not present in SIO. Considering that a hypothetical metallic state in SICO is dismissed by resistivity data, this result indicates that the insulating gap in SICO must be below $\sim 0.02$ eV, placing this material very close to a quantum critical point. The $|1/q|$ parameter tends to increase on warming, with anomalies at $T \sim 100$ K and $\sim T_{\text N}$. In particular, the anomaly at $T \sim 100$ K suggests there is another physically meaningful temperature below $T_{\text N}$ for SICO, as previously inferred for SIO \cite{Chikara2009,Ge2011,Bhatti_2014,Bhatti2019,Ye2020}. The reduced magnetization below $\sim 100$ K also supports this conclusion [see Fig. \ref{Mag}(a)].

Figure \ref{Raman2}(c) shows the temperature dependence of the frequency of the $M_3$ mode for SIO and SICO fitted with Lorentzian lineshapes ($|1/q|=0)$, in comparison to the expected anharmonic behavior $\Delta \omega_{\text {anh}}(T) = A [1+2/(\mathrm{e}^{\hbar \omega_0/2 k_{\text B} T}-1)]$ \cite{Balkanski1983}. Both compounds show anomalous shifts of this mode below $T_{\text N}$. This result is ascribed to the pseudospin-phonon coupling effect \cite{Baltensperger1968,Granado1999,Samanta2019}. The anomalous hardenings after subtraction of the anharmonic term are shown in Fig. \ref{Raman2}(d). The magnitude of the pseudospin-phonon anomaly is not reduced for SICO in comparison to SIO, indicating that the microscopic magnetic network of Ir pseudospins is still well preserved in SICO. The other $M_1-M_4$ phonon parameters are shown in the Supplementary Figure S7.

\section{Discussion}

We propose the schematic energy diagram displayed in Fig. \ref{struc}(c) to explain our experimental results. The quantum critical point occurs as the localized Co$^{3+}$ acceptor level matches the top of the lower Hubbard band at $E_{\text F}$. At low temperatures, the Co$^{3+}$ acceptor level is empty, thus the nominal oxidation state is Co$^{4+}$ and no charge carrier is present. Upon warming, such impurity level becomes gradually occupied by the electrons originated from the lower Hubbard band due to the thermal energy, shifting the Co oxidation state towards the Co$^{3+}$ side and hole-doping the lower Hubbard band. The density of states of the lower Hubbard band nearby $E_{\text F}$ is
\begin{equation}
n(E')= (1 / 4 \pi^2) (2 m^{*} / \hbar^2)^{3/2} \sqrt{E'},
\end{equation}
where $E' \equiv |E-E_{\text F}|$ and $m^{*}$ is the lower Hubbard band effective mass \cite{Ashcroft1976}. The number of thermally-activated holes per unit volume at low temperatures is
\begin{equation}
N(T)= \int_0^{\infty} n(E') \mathrm{e}^{-E'/k_{\text B} T} dE'=(1 / 4 \pi^2) (2 m^{*} / \hbar^2)^{3/2} \Gamma(3/2) (k_{\text B} T)^{3/2},
\end{equation}
where $\Gamma(u)$ is the gamma function. These carriers have mean energy $\bar{E}(T)=3k_{\text B} T/2$. We now treat these thermally-activated holes semiclassically with a renormalized mass $m_{\text {ren}}$ that is not necessarily equal to $m^{*}$. The time between collisions is
\begin{equation}
\tau = D/ \bar{v}= D \sqrt{m_{\text {ren}}/2 \bar{E}(T)}= D \sqrt{
m_{\text {ren}}/3k_{\text B} T},
\end{equation}
where $D$ is the carrier mean free path. From the Drude relation $\sigma(T) = q^2 N(T) \tau /m_{\text {ren}}$, where $q$ is the electron charge, we obtain
\begin{equation}
\sigma(T) = (\Gamma(3/2)/\sqrt{6} \pi^2)[q^2 (m^{*})^{3/2} D / \hbar^3 (m_{\text {ren}})^{1/2}] k_{\text B} T.
\end{equation}
Thus, this simple model is able to reproduce the observed power-law behavior for the electrical conductivity $\sigma(T)=A T^x$ with the correct exponent $x=1$. Also, the experimental coefficient $A=0.315$   $\Omega^{-1}$m$^{-1}$K$^{-1}$ [see Fig. \ref{rho}(b)] can be used to obtain further information on 
\begin{equation}
m_{\text {ren}}=(\Gamma^2(3/2)/6 \pi^4) (q^4 k_{\text B}^2/ \hbar^6) [(m^{*})^3 D^2 / A^2].
\end{equation}
Using $D \sim 8$ \AA\ (see above) and considering that $m^{*} > m_{\text e}$, we obtain $m_{\text {ren}} \gtrsim 700$ $m_{\text e}$, where $m_{\text e}$ is the free electron mass ! Such a huge estimated renormalized mass is indicative of heavy-fermion physics. Indeed, the energy diagram displayed in Fig. \ref{struc}(c), with a localized impurity level near $E_{\text F}$, suggests an analogy with heavy-fermion materials with localized $4f$ levels near $E_{\text F}$. The important difference here is that the ``conduction electrons'' are actually thermally-excited QCF. The absence of charge carriers as $T \rightarrow 0$ means there is no Sommerfeld coefficient $\gamma$ for the specific heat $c_{\text P}(T)$ associated with the QCF captured by our model. Rather, the low-temperature $c_{\text P}(T)$ from the QCF should actually scale with $d [N(T) \bar{E}(T)] / d T \propto T^{3/2}$, possibly mixing with magnon contributions. These considerations imply that the linear $\sigma(T)$ curve presented here is likely the most direct macroscopic manifestation of the QCF in SICO.

The observation of QCF in SICO, with similar $T_{\text N} \sim 240$ K and pseudospin-phonon coupling strength in comparison to SIO, demonstrates that the quantum critical points associated with the charge transport and magnetic transitions are distinct in these iridates. This is reasonable once a non-zero density of charge carriers is presumably necessary to destroy the magnetic ordering at $T=0$ K. The thermally-activated holes possess charge and spin-orbital degrees of freedom, and this system offers an opportunity to access the resulting QCF in the magnetically ordered regime.

\section{Conclusions}

In summary, a detailed analysis of our electrical resistivity and Raman scattering data uncovers the presence of QCF in SICO. The inferred quantum phase transition is related to the crossing of the Co$^{3+}$ acceptor level with the top of the lower $J_{\text {eff}}=1/2$ band at $E_{\text F}$. The linear behavior observed in $\sigma(T)$ is captured by a Drude-like semiclassical model that indicates a large renormalized electron mass for the thermally-activated charge carriers. 

\section{Methods}

\subsection{Sample synthesis}
The Sr$_2$IrO$_4$ (SIO) and Sr$_2$Ir$_{0.95}$Co$_{0.05}$O$_4$ (SICO) polycrystalline samples were synthesized by a standard solid-state reaction method employing high-purity SrCO$_3$, IrO$_2$ and CoO as described in Ref. \cite{Samanta2018}. Since the reagents are mixed in stoichiometric proportions, this method leaves minimum margin for compositional Sr/Ir/Co variations, which is critical for the conclusions of this work. Also, the large absolute resistivity values at low temperatures of SIO, and especially the large low-$T$/high-$T$ ratio $\rho(50$ K)/$\rho(300$ K) $\sim 10^4$ [see Fig. \ref{rho}(a)] compares to the highest reported values in the literature \cite{Klein2008}, unfavoring the possibility of any relevant doping due to off-stoichiometric oxygen content in this sample. This conclusion is likely extensible to SICO, considering that both samples were synthesized by the same procedure.

\subsection{X-ray diffraction and energy-dispersive X-ray spectroscopy}
Laboratory X-ray diffraction data were taken in a Bruker D2 diffractometer equipped with a linear detector, employing Cu $K \alpha$ radiation ($\lambda=1.542$ \AA). Rietveld refinements were performed with the  GSAS-II suite \cite{Toby2013}. The refined Co occupancy at the Ir site is 4.2(7) \% for SICO, close to the 5 \% nominal value. The other refined crystallographic parameters are given in the Supplementary Table S1. Energy-dispersive X-ray spectroscopy measurements were performed on two different spots of SICO using a scanning electron microscope FEI Nanolab 200 and confirm a homogeneous Co concentration of 5(1) \%. Synchrotron X-ray powder diffraction measurements at ambient temperature were also carried out at the XDS beamline of the Brazilian Synchrotron Light Laboratory (LNLS) \cite{Lima2016} with $\lambda = 0.61986$ \AA\, employing the experimental setup described in Ref. \onlinecite{Samanta2020}. The structure illustrations in Fig. \ref{struc} were produced with the aid of the program VESTA \cite{Momma2011}.

\subsection{Electrical and magnetic measurements}
Electrical resistivity and magnetoresistance were measured using a standard four-point technique with a {\it dc} bridge in an applied magnetic field up to $H=90$ kOe in a commercial system.  The {\it dc}-magnetization measurements were performed with a Superconducting Quantum Interference Device magnetometer (SQUID) and a SQUID with a Vibrating Sample Magnetometer (SQUID-VSM).

\subsection{Raman scattering}
Raman scattering experiments were performed in a quasi-backscattering geometry using the 488.0 nm line of Ar$^{+}$ laser with focus spot size $\sim 50$ $\mu$m. The laser power was kept below $20$ mW. A Jobin-Yvon T64000 triple-grating spectrometer with 1800 mm$^{-1}$ gratings was employed. A $L$N$_2$-cooled multichannel charge-coupled device was used to collect and process the scattered data. The sample was mounted on a cold finger of closed-cycle He cryostat with base temperature 20 K. 

\section{Acknowledgements}

We would like to acknowledge M. A. Avila for providing assistance in the SQUID-VSM measurements. This work was supported by Fapesp Grants 2016/00756-6, 2017/10581-1, 2018/20142-8, 2018/11364-7, and 2018/18653-4, as well as CNPq Grants 308607/2018-0, 304496/2017-0, and 409504/2018-1, Brazil. LNLS is acknowledged for the concession of beamtime. The authors also acknowledge the Center for Semiconducting Compounds and Nanotechnologies (CCS) at UNICAMP for providing the equipment and technical support for the energy-dispersive X-ray spectroscopy measurements.

\section{Data availability}
The authors declare that the main data supporting the findings of this study are available within the article and its Supplementary Information file. Extra data are available from the corresponding author upon reasonable request.

\section{Contributions}
K.S., J.C.S., P.G.P. and E.G. conceived the experiments. Experimental data were taken by K.S., J.C.S., D.R. and A.I.M. under supervision of P.G.P. and E.G. Data analysis was performed by K.S., J.C.S., A.I.M. and E.G. The model was developed by E.G. The paper was written by K.S. and E.G. with input from all authors.

\section{Competing interests}
The authors declare no competing interests.

\newpage

\begin{figure}[t]
\includegraphics[width=0.6\textwidth]{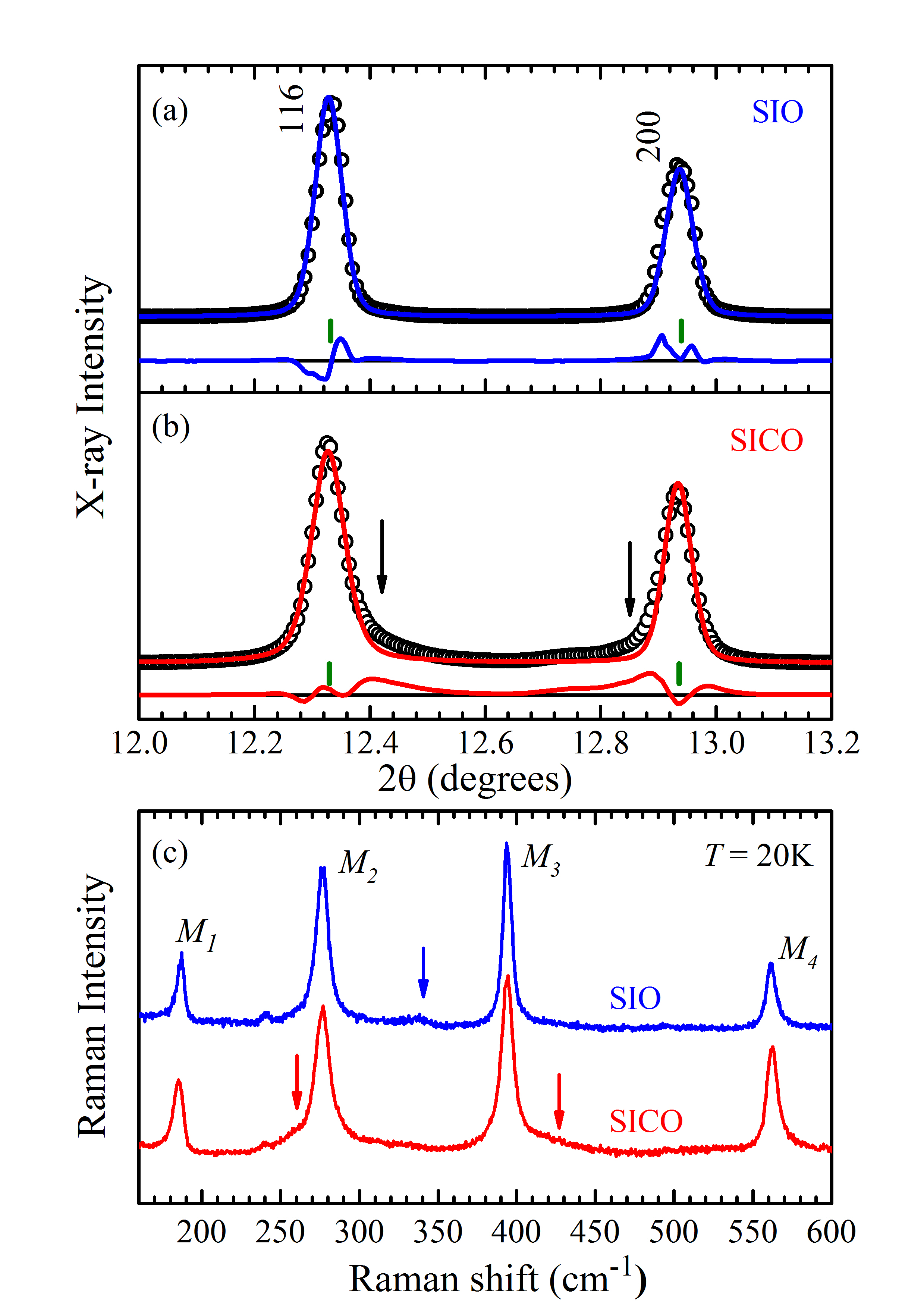}
\caption{\label{Raman1} {\bf Structural characterization.} Selected portion of the X-ray diffraction profiles of Sr$_2$IrO$_4$ (SIO, a) and Sr$_2$Ir$_{0.95}$Co$_{0.05}$O$_4$ (SICO, b) at room temperature ($\lambda=0.61986$ \AA), covering the 116 and 200 Bragg reflections marked by the short vertical bars. The open symbols are experimental data and the solid lines are calculated profiles after a Rietveld refinement using pseudo-Voigt lineshapes for the individual Bragg peaks. The difference profiles are shown at the bottom of each figure. The vertical arrows in (b) highlight the asymmetric lineshape broadening. (c) Unpolarized Raman scattering spectra of the two investigated samples at $T=20$ K. The vertical arrows highlight the presence of a peak at $\sim 340$ cm$^{-1}$ for SIO that is indicative of a stoichiometric sample \cite{Sung2016,Glamazda2014}, and extra features at $\sim 260$ and $\sim 420$ cm$^{-1}$ for SICO that are associated with weak disorder.}
\end{figure}
 
\begin{figure}[t]
	\includegraphics[width=0.9\textwidth]{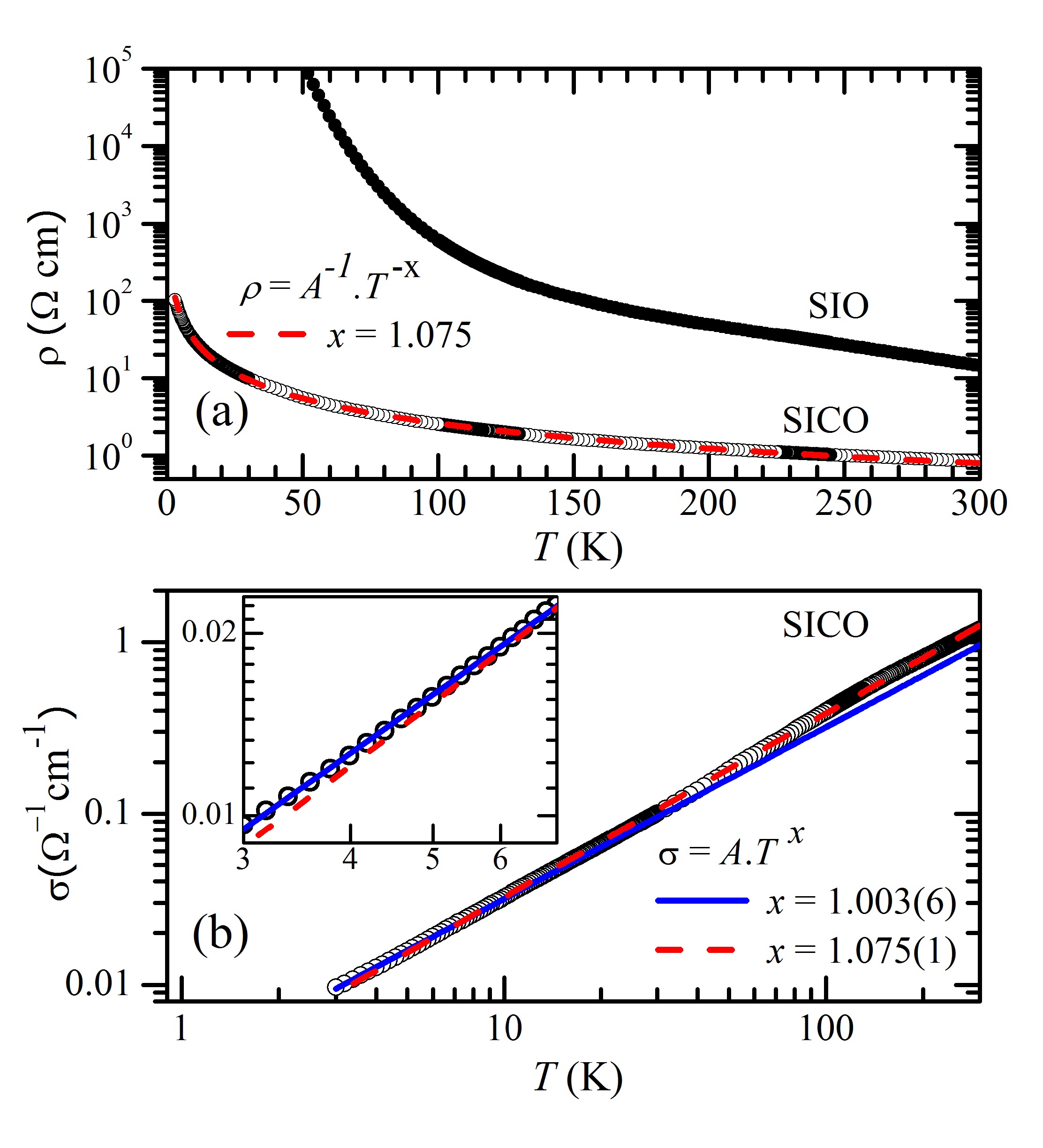}
\caption{\label{rho} {\bf Transport data at zero magnetic field.} (a) Temperature dependence of the electrical resistivity $\rho(T)$ of Sr$_2$IrO$_4$ (SIO) and Sr$_2$Ir$_{0.95}$Co$_{0.05}$O$_4$ (SICO). The dashed line is the fitted power-law behavior $\rho(T) = A^{-1} \cdot T^{-x}$, where $A$ and $x=1.075(1)$ are fitting constants, revealing that SICO is at the onset of a metal-insulator quantum phase transition. (b) Electrical conductivity $\sigma(T)$ in a double-log scale. The inset is a zoom out at low temperatures. The dashed and solid lines correspond to fits to $\sigma(T) = A \cdot T^{x}$ with $x=1.075(1)$ and $x=1.003(6)$, respectively, where the latter exponent was obtained from fitting low-temperature data only ($T<6$ K). Errors in parentheses are statistical only and represent one standard deviation obtained from the corresponding fit.}
\end{figure}

\begin{figure}[t]
\includegraphics[width=1.0\textwidth]{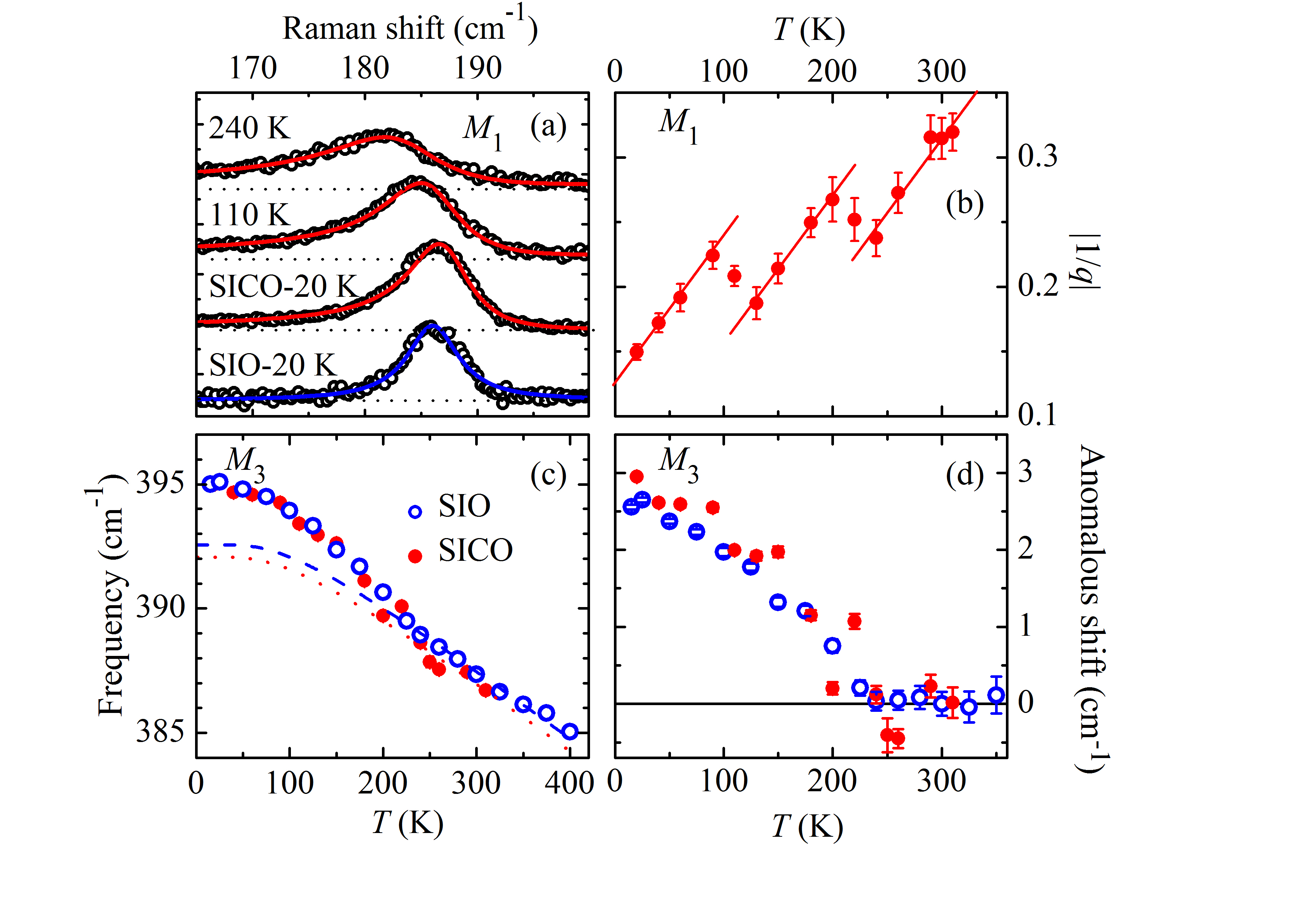}
\caption{\label{Raman2} {\bf Temperature-dependent Raman scattering.} (a) Raman spectrum of Sr$_2$Ir$_{0.95}$Co$_{0.05}$O$_4$ (SICO) at selected temperatures and Sr$_2$IrO$_4$ (SIO) at 20 K in the vicinity of the $M_1$ peak [symbols, see also Fig. \ref{Raman1}(c)]. The solid lines are fits to an asymmetric Fano lineshape for SICO and a symmetric Lorentzian lineshape for SIO. Data were vertically translated for clarity. The horizontal dotted lines represent the baseline for each spectrum. (b) Modulus of the Fano asymmetry parameter $|1/q|$ of the $M_1$ peak of SICO. The straight lines with equal angular coefficients are guides to the eye. (c) Frequency of the $M_3$ mode of SIO (open circles) and SICO (closed circles). The dashed and dotted lines indicate the expected anharmonic behavior for SICO and SIO, respectively \cite{Balkanski1983}. (d) Anomalous $M_3$ phonon shift, defined as the difference between the observed frequencies and the anharmonic contributions shown in (c). The errorbars in (b) and (d) are standard deviations extracted from the fits, which are smaller than the symbol sizes in (c).}
\end{figure}

\newpage

\newpage

\setcounter{figure}{0}  
\renewcommand{\thefigure}{S\arabic{figure}}
\renewcommand{\thetable}{S\arabic{table}}

\begin{table}
\caption{\label{strucpar} Refined crystallographic parameters for SIO and SICO at room temperature, using the data displayed in Fig. \ref{XRD}. The employed space group was $I4_{1}/\text{acd}$. The atomic positions are the following: Sr, (0,${1}\over{4}$,$z$); Ir/Co, (0,${1}\over{4}$,${3}\over{8}$); O1, (0,${1}\over{4}$,$z$); O2, ($x$,${1}\over{4}$ $+x$,${1}\over{8}$). Errors in parentheses are statistical only and represent one standard deviation.}
	
\begin{ruledtabular}
\begin{tabular}{ccc}
  & SIO & SICO \\
\hline
\multicolumn{3}{c}{Lattice constants}  \\
$a$ (\AA) & 5.49634(9)	& 5.49508(14) \\
$c$ (\AA) & 25.7923(5) &  25.7667(7) \\
$V$ (\AA$^3$) & 779.18(3)  & 778.05(5) \\
\hline
\multicolumn{3}{c}{Positional and thermal parameters}  \\
$U_{\text{iso}}^{\text{global}}$ (\AA$^2$) & 0.0449(4) & 0.0439(6) \\
Sr: $z$ & 0.55089(7) & 0.55085(12) \\
Ir/Co: frac & 1.005(3)/0 & 0.958(7)/0.042(7) \\
O1: $z$ & 0.4560(3) & 0.4524(5) \\
O2: $x$ & 0.1959(10) & 0.198(2) \\
\hline
\multicolumn{3}{c}{Bond distances and angles}  \\
Ir-O1 ($\times 2$, \AA) & 2.089(8) & 1.994(13) \\
Ir-O2 ($\times 4$, \AA) & 1.988(5) & 1.984(9) \\
Ir-O2-Ir (deg) & 155.6(3) & 156.7(4) \\
\hline
$R_w$ & 8.32 \% & 12.68 \% \\
 \\
\end{tabular}
\end{ruledtabular}
\end{table}

\begin{figure}[t]
\includegraphics[width=1\textwidth]{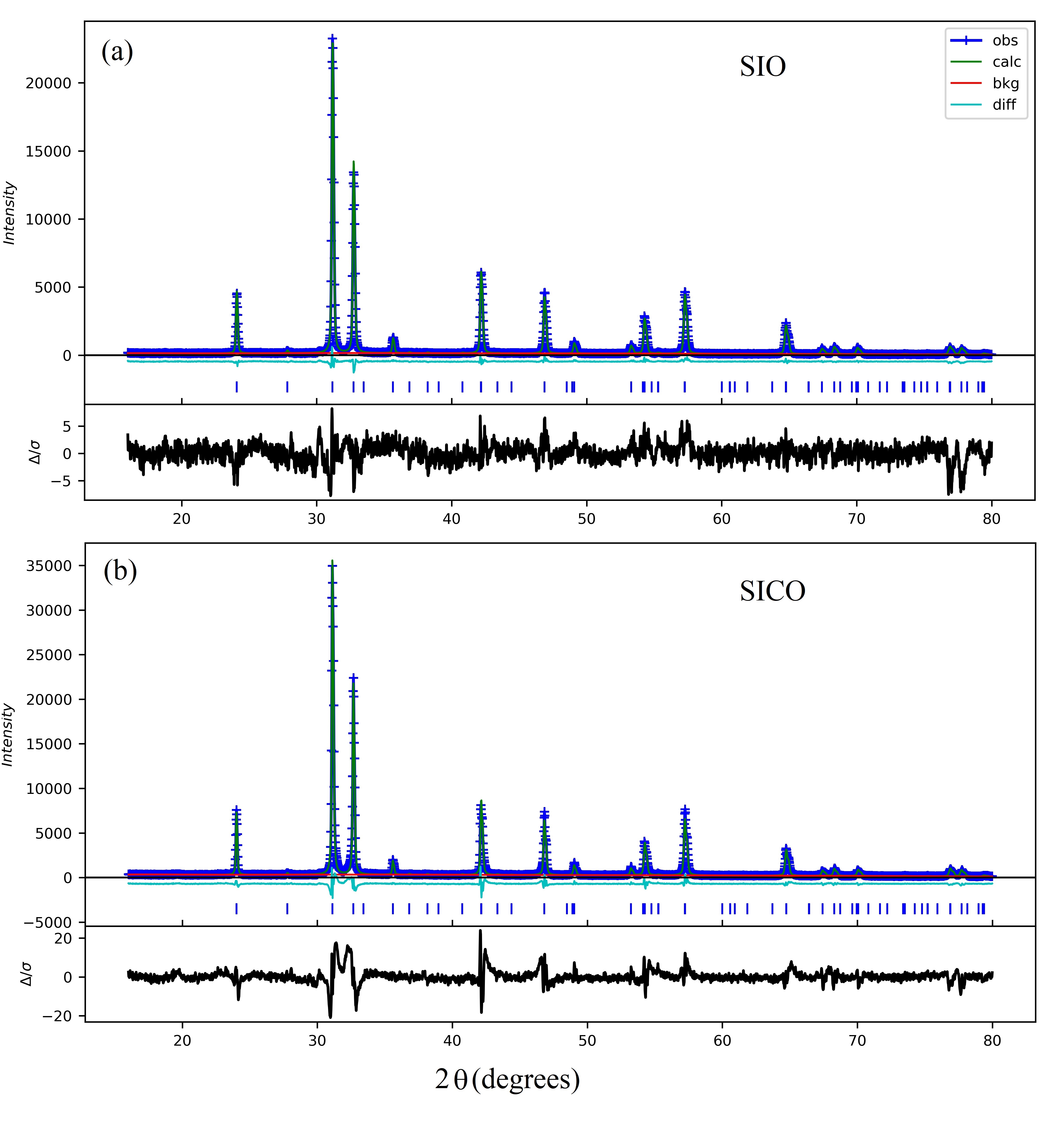}
\caption{\label{XRD} Laboratory X-ray diffraction profiles of Sr$_2$IrO$_4$ (SIO, a) and Sr$_2$Ir$_{0.95}$Co$_{0.05}$O$_4$ (SICO, b) at room temperature and $\lambda=1.542$ \AA\ taken in Bragg-Brentano geometry. The symbols are experimental data and the green solid lines are calculated profiles after a Rietveld fit using pseudo-Voigt lineshapes for the individual Bragg peaks. The short vertical bars mark the Bragg reflection positions. The bottom of each figure displays the difference between observed and calculated intensities $\Delta$ divided by the statistical standard variation $\sigma$ for each data point. Minor unidentified impurity peaks are found below the 0.2 \%\ intensity level with respect to the strongest peak in both samples and are not visible in this scale.}
\end{figure}

\begin{figure}[t]
\includegraphics[width=1\textwidth]{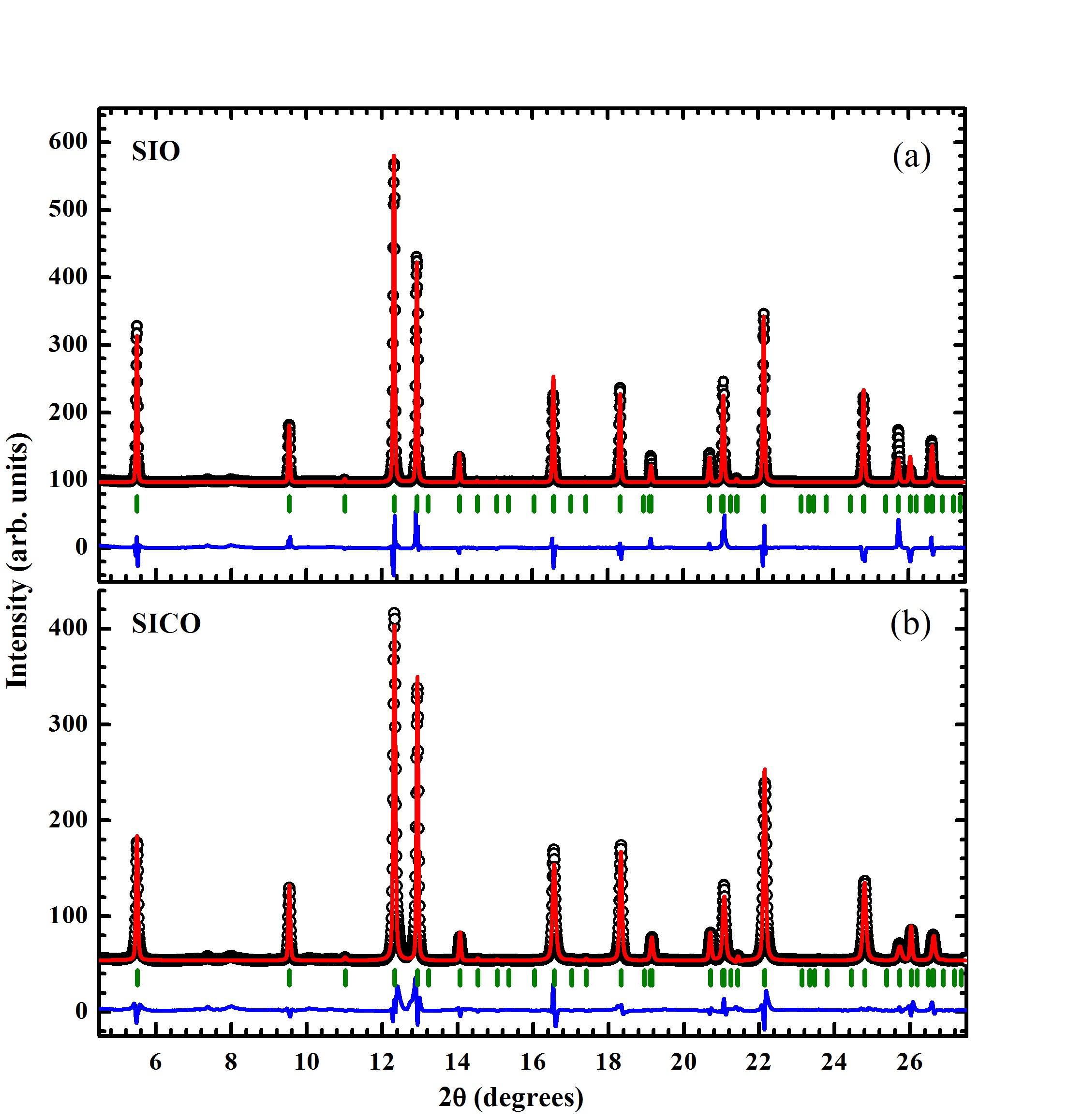}
\caption{\label{SXRD} Synchrotron X-ray diffraction profiles of SIO (a) and SICO (b) at ambient conditions with $\lambda=0.61986$ \AA\ taken in transmission geometry, using a kapton adhesive tape as sample support that contributes to broad structures in the $\sim 1$ \%\ intensity level. The open symbols are experimental data and the solid lines are calculated profiles after a Rietveld fit using pseudo-Voigt lineshapes for the individual Bragg peaks. The short vertical bars mark the Bragg reflection positions. The difference profiles are shown at the bottom of each figure.}
\end{figure}

%\section{Resistivity and magnetoresistance}

\begin{figure}[t]
\includegraphics[width=1\textwidth]{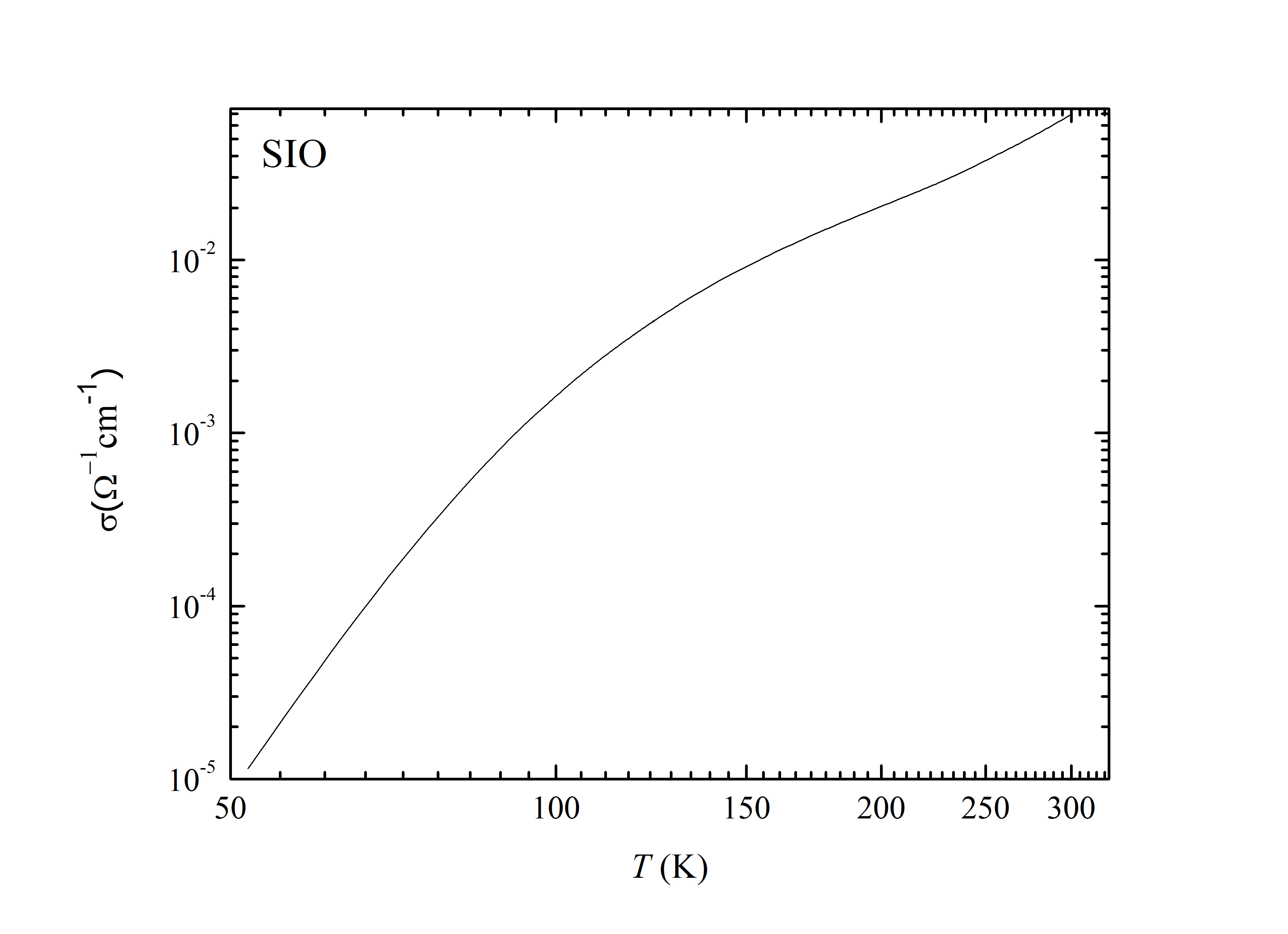}
\caption{\label{rhoSIO} $\sigma(T)$ for SIO in a log-log scale, showing that the pure sample does not obey a power-law behavior in constrast to SICO.}
\end{figure}

\begin{figure}[t]
\includegraphics[width=1\textwidth]{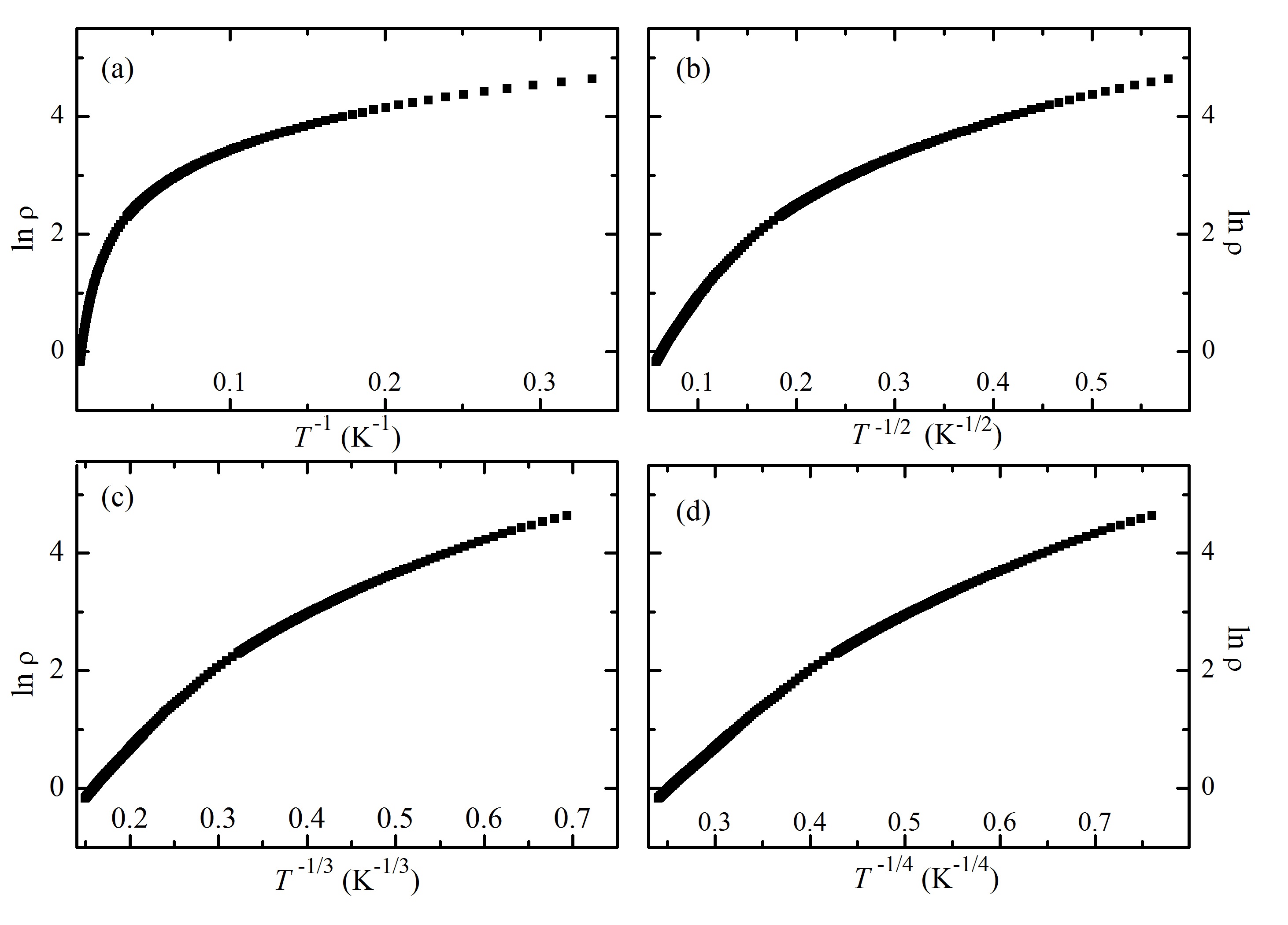}
\caption{\label{rho} ln$(\rho)$ as a function of $T^{-1}$ (a), $T^{-1/2}$ (b),  $T^{-1/3}$ (c) and $T^{-1/4}$ (d) for SICO. This analysis rules out an exponential behavior $\sigma(T) \propto \mathrm{exp}[-(T_0/T)^{\alpha}]$ with $1/4 \leq \alpha \leq 1$ for this sample.}
\end{figure}

\begin{figure}[t]
\includegraphics[width=1\textwidth]{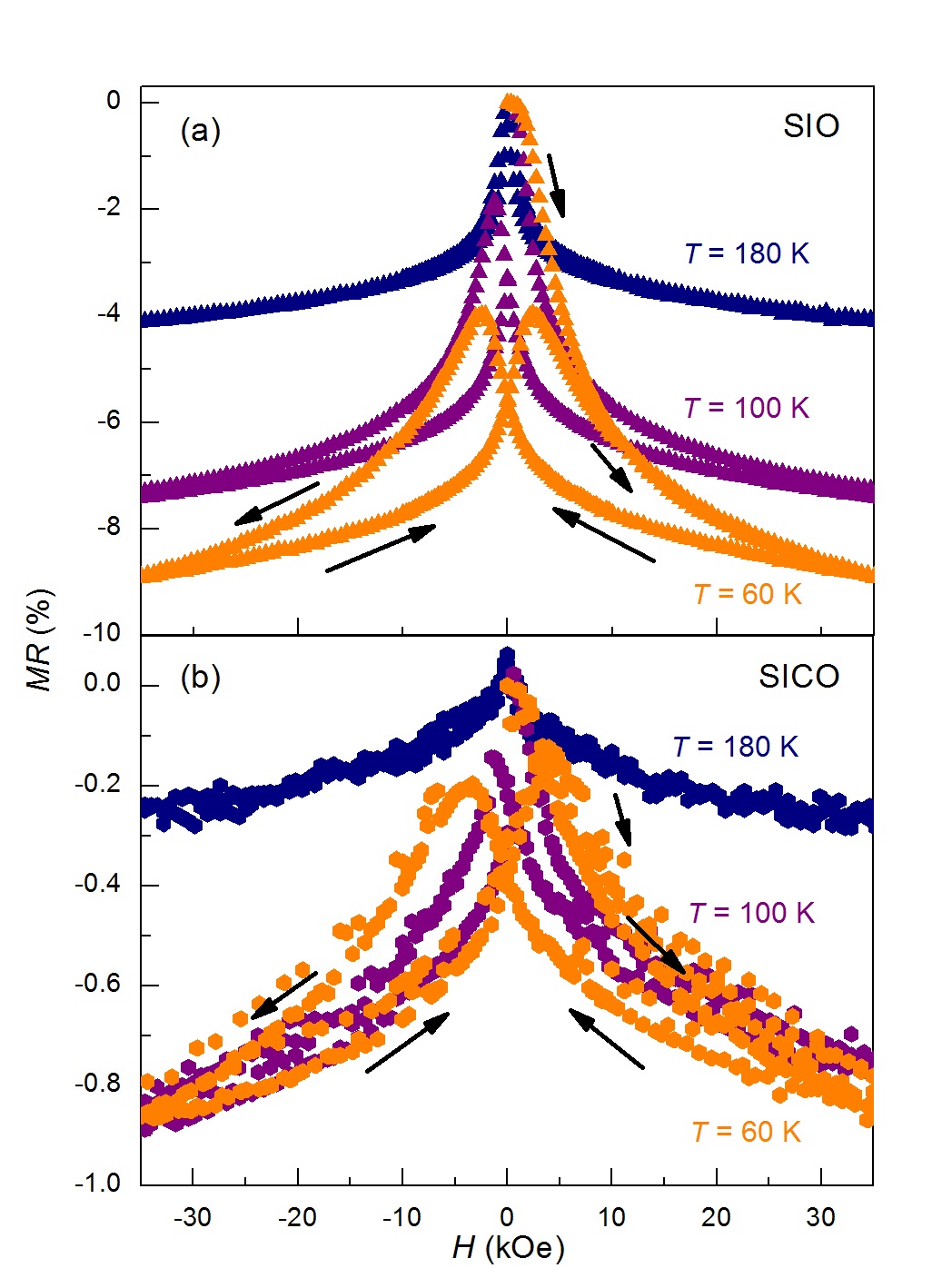}
\caption{\label{MR} Magnetoresistance (MR) of SIO (a) and SICO (b) at 60, 100 and 180 K.}
\end{figure}

\newpage

%\section{Raman scattering}

\begin{figure}[t]
\includegraphics[width=1\textwidth]{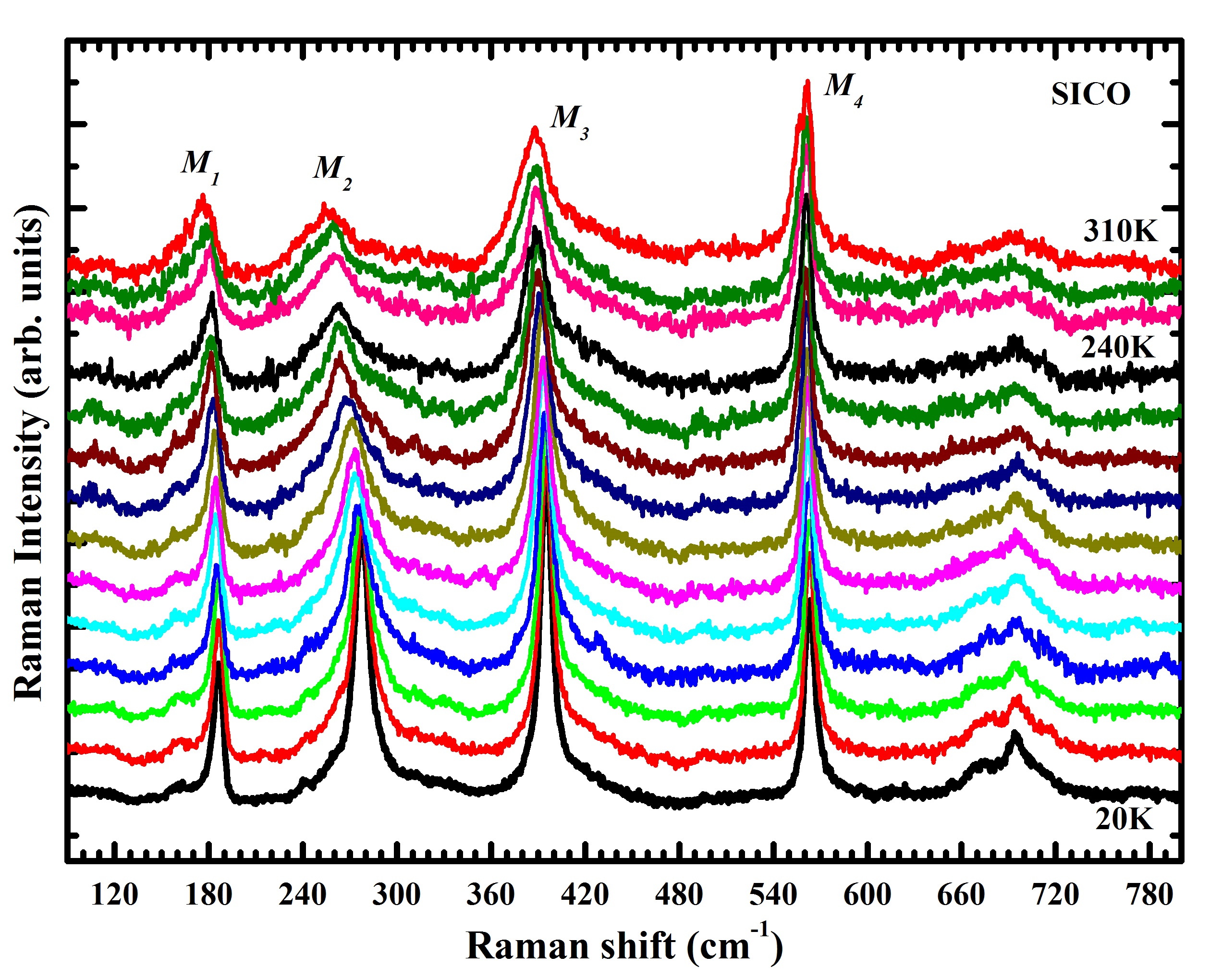}
\caption{\label{Ramanspectra} Unpolarized Raman spectra of SICO at several temperatures. The spectra are vertically translated for clarity.}
\end{figure}

\begin{figure}[t]
\includegraphics[width=1\textwidth]{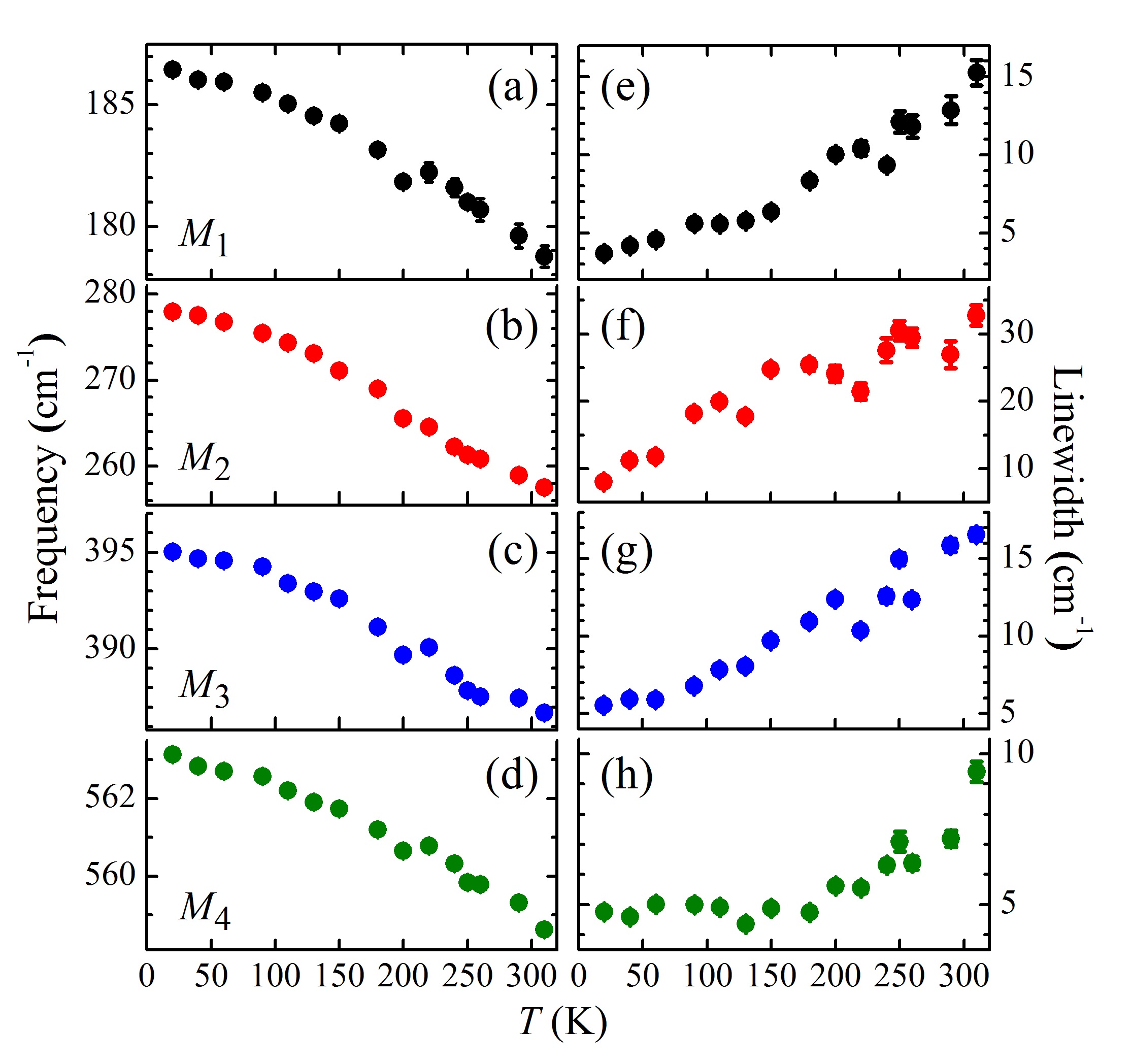}
\caption{\label{Ramanpar} Temperature dependence of the frequency (a-d) and linewidth (e-h) of the main Raman modes $M_1-M_4$ of SICO. The errorbars are standard deviations extracted from the fits, which are smaller than the symbol sizes when not visible.}
\end{figure}

\section{References}

\bibliographystyle{naturemag}
\bibliography{bibliography.bib}

\end{document}